\documentclass[aps,pra,10pt,a4paper,twocolumn,showpacs]{revtex4-1}

\pdfoutput=1
\usepackage{hyperref}
\usepackage{graphicx}
\usepackage{amsmath}
\usepackage{epsfig}
\usepackage[latin1]{inputenc}
\usepackage{amsfonts}
\usepackage{amssymb}

\newcommand{\hX}{\hat{X}}

\newcommand{\hP}{\hat{P}}

\newcommand{\lex}{\left<}
\newcommand{\rex}{\right>}

\date{\today}

\begin{document}

\title{Continuous variable entanglement distillation of Non-Gaussian Mixed States}

\author{Ruifang Dong$^{1,2,*}$, Mikael Lassen$^{1,2}$, Joel Heersink$^{1,3}$, Christoph Marquardt$^{1,3}$, Radim Filip$^4$, Gerd Leuchs$^{1,3}$ and Ulrik L. Andersen$^2$}

\affiliation{
$^{1}$Max Planck Institute for the Science of Light, Günther-Scharowsky-Str. 1/Bau 24, 91058 Erlangen, Germany\\
$^{2}$Department of Physics, Technical University of Denmark, Building 309, 2800 Lyngby, Denmark\\
$^{3}$Institute of Optics, Information and Photonics, Friedrich-Alexander-University Erlangen-Nuremberg, Str. 7/B2, 91058 Erlangen, Germany\\
$^{4}$Department of Optics, Palack\'y University, 17. listopadu 50, 77200 Olomouc, Czech Republic\\
$^{*}$rdon@fysik.dtu.dk}

\begin{abstract}
Many different quantum information communication protocols such as teleportation, dense coding and entanglement based quantum key distribution are based on the faithful transmission of entanglement between distant location in an optical network. The distribution of entanglement in such a network is however hampered by loss and noise that is inherent in all practical quantum channels. Thus, to enable faithful transmission one must resort to the protocol of entanglement distillation. In this paper we present a detailed theoretical analysis and an experimental realization of continuous variable entanglement distillation in a channel that is inflicted by different kinds of non-Gaussian noise. The continuous variable entangled states are generated by exploiting the third order non-linearity in optical fibers, and the states are sent through a free-space laboratory channel in which the losses are altered to simulate a free-space atmospheric channel with varying losses. We use linear optical components, homodyne measurements and classical communication to distill the entanglement, and we find that by using this method the entanglement can be probabilistically increased for some specific non-Gaussian noise channels.

\end{abstract}

\pacs{42.50.Lc,42.50.Dv,42.81.Dp,42.65.Dr}

\maketitle

\section{Introduction}

Quantum communication is a promising platform for sending secret
messages with absolute security and developing new low energy
optical communication systems~\cite{gisin02.rmp}. Such quantum communication protocols
require the faithful transmission of pure quantum states over very long distances. Heretofore,
significant experimental progress has been achieved in free space
and fiber based quantum cryptography where communication over more than 100~km have been demonstrated~\cite{dixon08.oe, Manderbach07.prl, ursin07.natphy}. The
implementation of quantum communication systems over even larger
distances - as will be the case for transatlantic or deep space
communication - can be carried out by using quantum teleportation.
However, it requires that the two communicating parties share highly
entangled states. One is therefore faced with the technologically
difficult problem of distributing highly entangled states over long
distances. The most serious problem in such a transmission is the
unavoidable coupling with the environment which leads to losses and
decoherence of the entangled states.

Losses and decoherence can be overcome by the use of entanglement
distillation, which is the protocol of extracting from an ensemble of
less entangled states a subset of states with a higher degree of
entanglement~\cite{bennett96.prl}. Distillation is therefore a purifying protocol that selects highly entangled pure states from a mixture that is a result of noisy transmission. This protocol has been
experimentally demonstrated for qubit systems
exploiting a posteriori generated polarization entangled
states~\cite{kwiat01.nat, zhao01.pra, yamamoto03.nat, pan03.nat, zhao03.prl}. Common for these implementations of entanglement distillation is their relative experimental simplicity; only simple linear optical components such as beam splitters and phase shifters are used to recover the entanglement. This inherent simplicity of the distillation setups arises from the non-Gaussian nature of the Wigner function of the entangled states. It has however been proved that in case the Wigner functions of the entangled states are Gaussian, entanglement distillation cannot be performed by linear optical components, homodyne detection and classical communication~\cite{eisert02.prl, fiurasek02.prl, giedke02.pra}. This is a very important result since it tells us that standard continuous variable entanglement generated by e.g. a second-order or a small third-order non-linearity cannot be distilled by simple means as these kinds of entangled states are described by Gaussian Wigner functions.

Several avenues around the no-go distillation theorem have been proposed. The first idea to increase the amount of CV entanglement was put forward by Opatrn{\'y} et al.~\cite{opatrny00.pra}. They suggested to subtract a single photon from each of the modes of a two-mode squeezed state using weakly reflecting beam splitters and single photon counters, and thereby conditionally prepare a non-Gaussian state which eventually could increase the fidelity of CV quantum teleportation. This protocol was further elaborated upon by Cochrane et al.~\cite{Cochrane2002.pra} and Olivares et al.~\cite{Olivares2003.pra}. Other approaches relying on strong cross Kerr nonlinearities were suggested by Duan et al.~\cite{duan00b.prl,duan2000.pra} and Fiur\'{a}\v{s}ek et al.~\cite{fiurasek2003.pra}. The usage of such non-Gaussian operations results in non-Gaussian entangled states. To get back to the Gaussian regime, it has been suggested to use a Gaussification protocol based on simple linear optical components and vacuum projective measurements (which can be implemented by either avalanche photodiodes or homodyne detection)~\cite{browne03.pra}. Distillation including the Gaussification protocol was first considered for pure states by Browne et al.~\cite{browne03.pra} but later extended to the more relevant case of mixed states by Eisert et al.~\cite{eisert03.pra}.

Due to the experimental complexity of the above mentioned proposals, the experimental demonstration of Gaussian entanglement distillation has remained a challenge. A first step towards the demonstration of Gaussian entanglement distillation was presented in ref.~\cite{ourjoumtsev07.prl} where a modified version of the scheme by Opatrn{\'y} et al.~\cite{opatrny00.pra} was implemented: single photons were subtracted from one of the two modes of a Gaussian entangled state using a nonlocal joint measurement and as a result, an increase of entanglement was observed. Recently, the full scheme of Opartny et al. was demonstrated by Takahashi et al.~\cite{takahashi09.quant}. They observed a gain of entanglement by means of conditional local subtraction of a single photon or two photons from a two-mode Gaussian state. Furthermore they confirmed that two-photon subtraction also improves Gaussian-like entanglement.

In the work mentioned above, only Gaussian noise has been considered as for example associated with constant attenuation.
Gaussian noise is however not the only kind of noise occurring in information channels. E.g. if the magnitude or phase of the transmission coefficient of a channel is fluctuating, the resulting transmitted state is a non-Gaussian mixed state. Because of the non-Gaussianity of the transmitted state, the aforementioned no-go theorem does not apply and thus Gaussian transformations suffice to distill the state. Actually, such a non-Gaussian mixture of Gaussian states can be distilled and Gaussified using an approach~\cite{fiurasek07.pra} related to the one suggested in ref.~\cite{browne03.pra}. Alternatively, it is also possible to distill and Gaussify non-Gaussian states using a simpler single-copy scheme which is not relying on interference but is based on a weak measurement of the corrupted states and heralding of the remaining state~\cite{dong08.nat}. Such an approach has been also suggested for cat state purification~\cite{suzuki06.pra}, coherent state purification~\cite{wittmann08.pra} and squeezed state distillation~\cite{heersink06.prl}.

The distillation of Gaussian entanglement corrupted by non-Gaussian noise was recently experimentally demonstrated in two different laboratories. More specifically, it was demonstrated that by employing simple linear optical components, homodyne detection and feedforward, it is possible to extract more entanglement out of a less entangled state that has been affected by attenuation noise~\cite{dong08.nat} or phase noise~\cite{hage08.nat}.

In this paper we elaborate on the work in ref.~\cite{dong08.nat}, largely extending the theoretical discussion on the characterization of non-Gaussian entanglement and, on the experimental side, testing our distillation protocol in new attenuation channels.

The paper is organized as follows: In Section~\ref{theory}, the entanglement distillation protocol utilized in our experiment is fully discussed. In Section~\ref{experiment} the experimental setup for realization of the entanglement distillation is described, and the experimental results are shown and discussed in Section~\ref{Experimental results}.

\section{The entanglement distillation protocol}\label{theory}

The basic scheme of entanglement distillation is illustrated in Fig.~\ref{schematics}. The primary goal is to efficiently distribute bipartite entanglement between two sites in a communication network to be used for e.g. teleportation or quantum key distribution. Suppose the two-mode entangled state (also known as Einstein-Podolsky-Rosen state (EPR)) is produced at site A. One of the modes is kept at A's site while the second mode is sent through a noisy quantum channel. As a result of this noise, the entangled state will be corrupted and the entanglement is degraded. The idea is then to recover the entanglement using local operations at the two sites and classical communication between the sites. To enable distillation, it is however required to generate and subsequently distribute a large ensemble of highly entangled states. After transmission, the ensemble transforms into a set of less entangled states from which one can distill out a smaller set of higher entangled states.

A notable difference between our distillation approach and the schemes proposed in Refs.~\cite{browne03.pra,fiurasek07.pra} is that our procedure relies on single copies of distributed entangled states whereas the protocols in Ref.~\cite{browne03.pra,fiurasek07.pra} are based on at least two copies. The multi copy approach relies on very precise interference between the copies, thus rendering this protocol rather difficult. One disadvantage of the single copy approach is the fact that the entangled state is inevitably polluted with a small amount of vacuum noise in the distillation machine. This pollution can, however, be reduced if one is willing to trade it for a lower success rate.

Before describing the details of the experimental demonstration, we wish to address the question on how to evaluate the protocol. The entanglement after distillation must be appropriately evaluated and shown to be larger than the entanglement before distillation to ensure a successful demonstration. One way of verifying the success of distillation is to fully characterise the input and output states using quantum tomography and then subsequently calculate an entanglement monotone such as the logarithmic negativity. However, in the experiment presented in this paper (as well as many other experiments on continuous variable entanglement) we only measured the covariance matrix as such measurements are easier to implement. The question that we would like to address in the following is whether it is possible to verify the success of distillation based on the covariance matrix of a non-Gaussian state.

\subsection{Entanglement evaluation}\label{sec_entevaluation}

In order to quantify the performance of the distillation protocol, the amount of distillable entanglement before and after distillation ought to be computed. It is however not known how to quantify the degree of distillable entanglement of non-Gaussian mixed states~\cite{bennett96.prl, rains99.pra1, rains99.pra2}. Therefore, as an alternative to the quantification of the distillation protocol, one could try to estimate qualitatively whether distillation has taken place by comparing computable bounds on distillable entanglement before and after distillation. First we will have a closer look at such bounds.

\subsubsection{Upper and lower bounds on distillable entanglement}\label{sec_bounds}

\begin{figure}[h]
\includegraphics[width=0.48\textwidth]{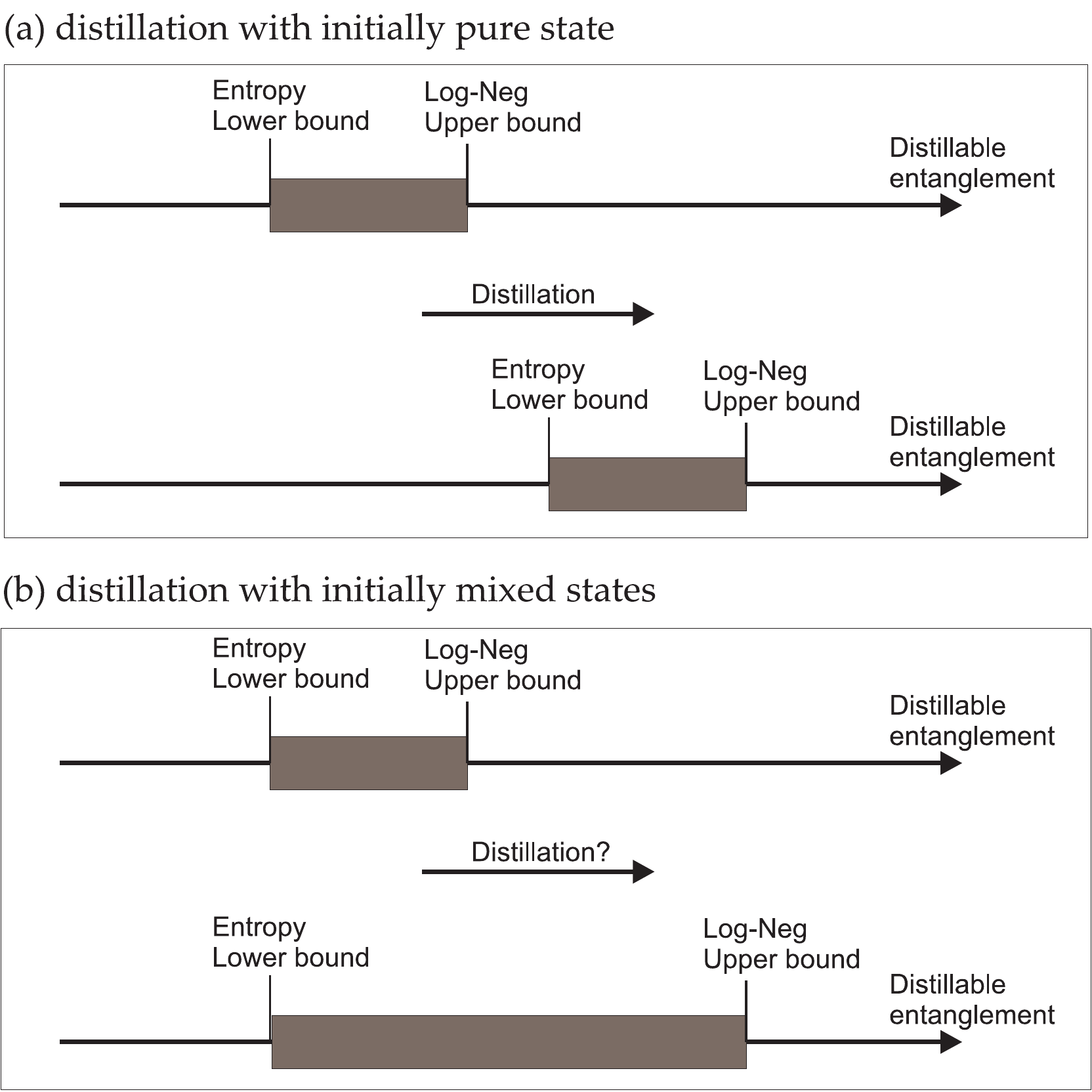}
\caption{Schematic demonstration of entanglement distillation of non-Gaussian mixed states. In figure (a), the distillation with a pure state is illustrated via the shift of the entanglement interval composed by the upper and lower bounds on distillable entanglement before and after the distillation protocol. Figure (b) shows the distillation with mixed states, the lower bound of which does not manifest increase even for a small excess noise in the state.}
\label{bounds}
\end{figure}

Although it is unknown how to find the amount of distillable entanglement of non-Gaussian mixed states, we can easily find the upper and lower bounds by computing the logarithmic negativity and the conditional entropy, respectively~\cite{vidal02.pra,lower2,lower1}. These bounds can be found before and after distillation, and the success of the distillation protocol can be unambiguously proved by comparing these entanglement intervals: If the entanglement interval is shifted towards higher entanglement and is not overlapping with the interval before the distillation, the distillation has proved successful. In other words, distillation has been performed if the lower bound after the protocol is larger than the upper bound before. This is illustrated in Fig.~\ref{bounds}(a).

It has been proved that for any state, the {\em log-negativity},
\begin{equation}
LN(\rho) \equiv \log_2 \left(2 \mathcal{N} + 1 \right) = \log_2 \left\|\rho^{T_A}\right\|_1.	
\end{equation}
is an upper bound on the distillable entanglement; $E_D < LN(\rho)$ ~\cite{vidal02.pra}. Here $\rho$ is the density matrix of the state, $||\rho^{T_A}||$ is the trace norm of the partial transpose of the state with respect to subsystem A, and the {\em negativity} is defined as
\begin{equation}
\mathcal{N}(\rho) \equiv \frac{\left\|\rho^{T_A}\right\|_1 -1}{2}.	
\end{equation}
The {\em negativity} corresponds to the absolute value of the sum of the negative eigenvalues of $\rho^{T_A}$ and it vanishes for non-entangled states.

In our experiment we were not able to measure the density matrix and thus compute the exact value of the negativity. We therefore use another (more strict) upper bound that is experimentally easier to estimate. As the {\em negativity} is a convex function we have
\begin{equation}
\mathcal{N}(\sum_{i} p_i \rho_i) \leq \sum_{i} p_i \mathcal{N}(\rho_i).	
\label{N_convex}
\end{equation}
where $\rho_i$ denotes the $i$th hermitian component in the mixed state, and $p_i$ is the weight for the $i$th component with $p_i \geq 0$ and $\sum_{i} p_i =1$.
Using this result we can find an upper bound on the log-negativity for mixed states:
\begin{equation}
LN(\sum_{i} p_i \rho_i) \leq \log_2 \left( 1 + 2 \sum_{i} p_i \mathcal{N}(\rho_i)\right).	
\label{LN_convex}
\end{equation}
This upper bound for the log-negativity will later be used to compute an upper bound for the distillable entanglement.

Another entanglement monotone is the conditional entropy. In contrast to the log-negativity, the conditional entropy yields a lower bound on the distillable entanglement: $E_D>S(\tilde{\rho}_A)-S(\tilde{\rho})$ \cite{lower1, lower2}, where $\tilde{\rho}$ is the density matrix corresponding to Gaussian approximation of the state and $\tilde{\rho}_A$ is the reduced density matrix with respect to system A. The entropies of the states can be calculated from the covariance matrix, $\mathbf{CM}$, using
\begin{eqnarray}
S(\tilde{\rho}_A)&=& f(\det \mathbf{A}),\nonumber\\
S(\tilde{\rho})&=&\sum_{i}f(\mu_i),\nonumber\\
f(x)&=&\frac{x+1}{2}\log_2(\frac{x+1}{2})-\frac{x-1}{2}\log_2(\frac{x-1}{2}),
\end{eqnarray}
where
\begin{equation}
\mu_{1,2}=\sqrt{\frac{\gamma \pm \sqrt{\gamma^2-4\det \mathbf{CM}}}{2}},
\end{equation}
are the symplectic eigenvalues of the covariance density matrix and $\gamma=\det \mathbf{A} + \det \mathbf{B} + 2 \det \mathbf{C}$. Here $\mathbf{A}$, $\mathbf{B}$, and $\mathbf{C}$ are submatrices of the covariance matrix: $\mathbf{CM}=\{\mathbf{A},\mathbf{C};\mathbf{C}^T,\mathbf{B}\}$.

It is important to note that this lower bound is very sensitive to excess noise of the two-mode squeezed state. Even for a small amount of excess noise, the lower bound approaches zero and thus is not very useful. This is illustrated in Fig.\ref{bounds}(b) which shows the distillable entanglement intervals before and after distillation of a noisy entangled state. Although the distillation protocol might remove the non-Gaussian noise of the state, the Gaussian noise of the state persists, and thus the entropy (that is the lower bound on distillable entanglement) will remain very low even after distillation. This results in an overlap between the two entanglement intervals and thus the comparison of computable entanglement bounds fails to witness the action of distillation in terms of distillable entanglement.

\subsubsection{Logarithmic negativity}
In our experiment, the entangled states possess a large amount of Gaussian excess noise and thus the prescribed method is insufficient to prove the act of entanglement distillation using distillable entanglement as a measure. However, in certain cases we can use the logarithmic negativity as a measure to witness the act of entanglement distillation even though we only have access to the covariance matrix as we will explain in the following.

First we note that in general, the Gaussian logarithmic negativity is an insufficient measure of entanglement distillation of non-Gaussian states as this measure only yields an upper bound, and with upper bounds of LN both before and after distillation a conclusion cannot be drawn. However, if the state after distillation is perfectly Gaussified its Gaussian LN becomes the exact LN, and if this exact value of LN is larger than the upper bound of LN before distillation (computed from (eqn.~\ref{LN_convex})), one may successfully prove the action of entanglement distillation entirely from the covariances matrices. This condition will be used for some of the experiments presented in this paper. More specifically, we will use this approach for testing entanglement distillation in a binary transmission channel. For other transmission channels investigated in this paper, the state will not be perfectly Gaussified in the distillation process and the approach cannot be applied. For such cases, however, we will resort to evaluations of the Gaussian part of the state in terms of Gaussian entanglement.

\subsubsection{Gaussian entanglement}

In addition to an increase in distillable entanglement and logarithmic negativity, the protocol can also be evaluated in terms of its Gaussian entanglement. Although the Gaussian entanglement is not accounting for the entanglement of the entire state (but only considers the second moments), it is quite useful as it directly yields the amount of entanglement useful for Gaussian protocols, a prominent example being teleportation of Gaussian states.

In a Gaussian approximation, the state can be described by the covariance matrix $\mathbf{CM}$~\cite{walls94.book}. The logarithmic negativity ($LN$) can then be found as
\begin{equation}\label{eq_LN}
LN = -\log_2 \nu_{min}.	
\end{equation}
where $\nu_{min}$ is the smallest symplectic eigenvalue of the partial transposed covariance matrix. The symplectic eigenvalues can be calculated from the covariance matrix using
\begin{equation}\label{eq_nu}
\nu_{1, 2} = \sqrt{\frac{\delta \pm \sqrt{\delta^2 - 4 \det \mathbf{CM}}}{2}}	
\end{equation}
where $\delta = \det \mathbf{A} + \det \mathbf{B} - 2 \det \mathbf{C}$, $\mathbf{A}$, $\mathbf{B}$ and $\mathbf{C}$ represent the submatrices in the correlation matrix~\cite{vidal02.pra}. Then by finding the smallest eigenvalue of the covariance matrix and inserting it in eqn.~(\ref{eq_LN}), a measure of the Gaussian entanglement of the state can be found.

\subsection{Theory of our protocol}

\begin{figure}[ht]
\includegraphics[width=0.48\textwidth]{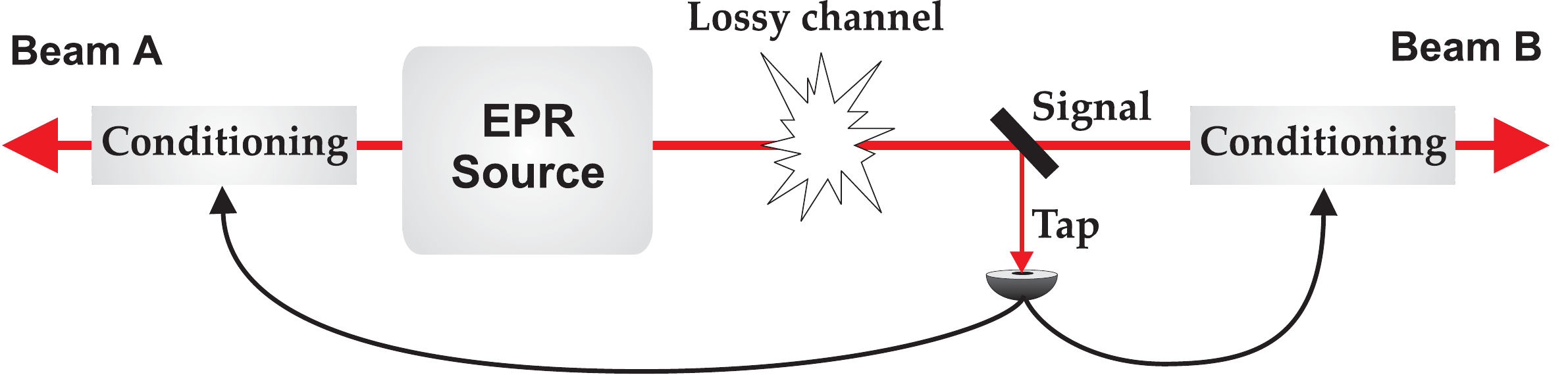}
\caption{Schematics of the entanglement distillation protocol. A weak measurement on beam B is diagnosing the state and
subsequently used to herald the highly entangled components of the state.}
\label{schematics}
\end{figure}

We now undertake our experimental setup a theoretical treatment in light of the results of the previous section.

The schematic of our protocol is shown in Fig.~\ref{schematics}. The two-mode squeezed or entangled state is produced by mixing two squeezed Gaussian states at a beam splitter. The squeezed states are assumed to be identical with variances $V_S$ and $V_A$ along the squeezed and anti-squeezed quadratures, respectively. The beam splitter has a transmittivity of $T_S$ and a reflectivity of $R_S =1-T_S$. One mode (beam A) from the entangled pair is given to Alice and the other part (beam B) is transmitted through a fading channel. The loss in the fading channel is characterized by the transmission factor $0\leq \eta(t)\leq 1$ which fluctuates randomly. The probability distribution of the fluctuating attenuation can be divided into $N$ different slots each associated with a sub-channel with a constant attenuation. The transmission of sub-channel $i$ is $\eta_i$ and it occurs with the probability $p_i$ so that $\sum^{N}_{i=1}p_i=1 $. For a particular $i$th sub-channel with transmission of $\eta_i$, the transmitted state is Gaussian and can be fully characterised by the covariance matrix $\mathbf{CM}_i$:
\begin{eqnarray}
\mathbf{CM}_i &=& \left(
\begin{array}{cc}
\mathbf{A}_i & \mathbf{C}_i \\
\mathbf{C}^T_i & \mathbf{B}_i
\end{array}
\right) ,\quad
\mathbf{A}_i =  \left(
\begin{array}{cc}
V_{AX,i} & 0 \\
0 & V_{BX,i}
\end{array}
\right), \nonumber\\
\mathbf{B} &=&  \left(
\begin{array}{cc}
V_{AP,i} & 0 \\
0 & V_{BP,i}
\end{array}
\right),\quad
\mathbf{C} =  \left(
\begin{array}{cc}
C_{X,i} & 0 \\
0 & C_{P,i}
\end{array}
\right).
\end{eqnarray}
where the elements are given by:
\begin{eqnarray}\label{mom1}
V_{AX,i}&=&T_S V_S+R_S V_A,\nonumber\\
V_{BX,i}&=&\eta_i(T_S V_A+R_S V_S)+(1-\eta_i),\nonumber\\
V_{AP,i}&=&T_S V_A+R_S V_S,\nonumber\\ V_{BP,i}&=&\eta_i(T_S V_S+R_S V_A)+(1-\eta_i),\nonumber\\
C_{X,i} &=&-C_{P,i}=\sqrt{\eta_i}\sqrt{ R_S T_S}(V_A-V_S).
\end{eqnarray}
Then according to eqn.~(\ref{LN_convex}) we can find an upper bound for the log-negativity of the state after transmission in the fluctuating channel (using$||\rho_i^T||=1/\nu_{min,i}$):
\begin{eqnarray}
LN(\sum_{i} p_i \rho_i) \leq \log_2 \sum_{i}(p_i/\nu_{min,i}).	
\end{eqnarray}
where $\nu_{min,i}$ corresponds to the smallest symplectic eigenvalue of the $i$th partial transposed covariance matrix. This means that the right hand side of this expression is also an upper bound on the distillable entanglement of the non-Gaussian noisy state. Therefore, to truly prove that the entanglement has increased, this bound must in principle be surpassed.

We now consider the Gaussian entanglement of our states using the Wigner function formalism. The Wigner function of the  total state and the $i$th state can be described as
\begin{eqnarray}\label{sum}
W(\mathbf{X},\mathbf{P})&=&\sum_{i}^{N} p_i W_i(\mathbf{X},\mathbf{P}),\nonumber\\
W_i(\mathbf{X},\mathbf{P})&=&\frac{\exp\left(-\mathbf{X} \mathbf{V}_{X,i}^{-1} \mathbf{X}^{T}-\mathbf{P} \mathbf{V}_{P,i}^{-1} \mathbf{P}^{T} \right)}{4\pi^{2}\sqrt{\mbox{det} \mathbf{V}_{X,i}\; \mbox{det} \mathbf{V}_{P,i}}},
\end{eqnarray}
where $\mathbf{X}=(x_{A},x_B)$ and $\mathbf{P}=(p_A,p_B)$. $\mathbf{V}_{X,i}$ and $\mathbf{V}_{P,i}$ are given by
\begin{eqnarray}
\mathbf{V}_{X,i} = \left(
\begin{array}{cc}
V_{AX,i} & C_{X,i} \\
C_{X,i} & V_{BX,i}
\end{array}
\right) ,\;
\mathbf{V}_{P,i} =  \left(
\begin{array}{cc}
V_{AP,i} & C_{P,i} \\
C_{P,i} & V_{BP,i}
\end{array}
\right).
\end{eqnarray}

From the Wigner function the second moments of the quadratures can be calculated through integration:
\begin{eqnarray}
\left\langle \hat{Z} \hat{Y} \right\rangle = \int dx_A dx_B dp_A dp_B z y W(x_{A},x_B,p_A,p_B)\nonumber\\
=\sum_i p_i \int dx_A dx_B dp_A dp_B z y W_i(x_{A},x_B,p_A,p_B),
\end{eqnarray}
where $\hat{Z},\hat{Y}=\hX_{A},\hX_B,\hP_A,\hP_B$. As the first moments of the vacuum squeezed states in both the quadratures are zero, the variances directly correspond to the second moments. Therefore, the elements of the total covariance matrix are simply the convex sum of the symmetrical moments (\ref{mom1}):
\begin{equation}
\langle \hat{Z} \hat{Y}\rangle=\sum_i p_i \langle \hat{Z} \hat{Y} \rangle_{i}.
\end{equation}
Since all the moments (\ref{mom1}) are just linear combinations of the transmission factors $\eta_i$ and $\sqrt{\eta_i}$, the covariance matrix of the mixed state has the following elements:
\begin{eqnarray}\label{mom2}
V_{AX}&=&T_S V_S+R_S V_A,\nonumber\\ V_{BX}&=&\lex\eta\rex(T_S V_A+R_S V_S)+(1-\lex\eta\rex),\nonumber\\
V_{AP}&=&T_S V_A+R_S V_S,\nonumber\\ V_{BP}&=&\lex\eta\rex(T_S V_S+R_S V_A)+(1-\lex\eta\rex),\nonumber\\
C_{X} &=& -C_{P}=\lex\sqrt{\eta}\rex\sqrt{R_S T_S}(V_A-V_S).
\end{eqnarray}
where the symbol $\lex \cdot \rex$ denotes averaging over the fluctuating attenuations. Comparing this set of equations with the set in (\ref{mom1}) associated with the second moments for the single sub-channels, we see that the attenuation coefficient  $\eta$ is replaced by the averaged attenuation $\lex\eta\rex$, and $\sqrt{\eta}$ is replaced by  $\lex\sqrt{\eta}\rex$. It is interesting to note that if the attenuation factor is constant (which means that the transmitted state will remain Gaussian) there will always be some, although small, amount of Gaussian entanglement left in the state. On the other hand, if the attenuation factor is statistically fluctuating as in our case, the Gaussian entanglement of the non-Gaussian state will rapidly degrade and eventually completely disappear.

To implement entanglement distillation, a part of the beam B is extracted by a tap beam splitter with transmittivity $T$. A single quadrature is measured (for example the amplitude quadrature, $\hX_t$) and based on the measurement outcome the remaining state is probabilistically heralded; it is either kept or discarded depending on whether the measurement outcome is above or below the threshold value $x_{th}$. The conditioned Wigner function of the output signal state after the distillation is
\begin{eqnarray}
\lefteqn{W_p(x_A,p_A,x'_B,p'_B)=}\nonumber\\
&&\int_{x_{th}}^\infty dx_t \int_{-\infty}^\infty dp_t \sum_{i=1}^N p_iW_i(x_A,p_A,x_B,p_B)W_0(x_v,p_v).\nonumber\\
&& \label{postmeasured_state1}
\end{eqnarray}
where $x_B=\sqrt{T}x'_B-\sqrt{1-T}x_t$, $p_B=\sqrt{T}p'_B+\sqrt{1-T}p_t$, $x_v=\sqrt{T}x_t-\sqrt{1-T}x'_B$ and $p_v=\sqrt{1-T}p_t+\sqrt{T}p'_B$, the Wigner function $W_0(x_v,p_v)$ represents the vacuum mode entering the asymmetric tap beam splitter. After integration, the Wigner function can be written as
\begin{eqnarray}
\lefteqn{W_p(x_A,p_A,x'_B,p'_B)=}\nonumber\\
&&\frac{1}{P_S}\sum_i p_i W'_{X,i}(x_A,x'_B;x_{th})\times W'_{P,i}(p_A,p'_B).
\end{eqnarray}
This is a product mixture of two non-Gaussian states which should be compared to the state before distillation which was a mixture of Gaussian states. $P_S$ is the total probability of success.  

The $\hX$ related elements of the covariance matrix can be calculated from this Wigner function directly by computing the symmetrically ordered moments:
\begin{eqnarray}
\langle X_A\rangle^{P}&=&\frac{\sum_i p_i \langle X'_A\rangle_i^{P}}{P_S}, \nonumber\\
\langle X'_B\rangle^{P}&=&\frac{\sum_i p_i \langle X'_B\rangle_i^{P}}{P_S},\nonumber\\
\langle X^{2}_A\rangle^{P}&=&\frac{\sum_i p_i \langle X_A^{'2}\rangle_i^{P}}{P_S},\nonumber\\
\langle X^{'2}_B\rangle^{P}&=& \frac{\sum_i p_i \langle X_B^{'2}\rangle_i^{P}}{P_S},\nonumber\\
\langle X_A X'_B\rangle^{P}&=&\frac{\sum_i p_i \langle X_A X'_B\rangle_i^{P}}{P_S}.
\end{eqnarray}
with
\begin{eqnarray}
\langle X_A\rangle_i^{P}&=&\frac{C_{X,i}\sqrt{R}}{\sqrt{2\pi V'_{DX,i}}}\exp\left(-\frac{x_{th}^2}{2 V'_{DX,i}} \right),\nonumber\\
\langle X'_B\rangle_i^{P}&=&\frac{\sqrt{TR}(V_{BXi}-1)}{\sqrt{2\pi V'_{DX,i} }}\exp\left(-\frac{x_{th}^2}{2 V'_{DX,i}} \right),\nonumber\\
\langle X^{2}_A\rangle_i^{P}&=&\frac{RC_{X,i}^{2} x_{th}}{\sqrt{2\pi V^{'3}_{DX,i}}}\exp\left(-\frac{x_{th}^2}{2 V'_{DX,i}} \right)+ \nonumber\\
&&\frac{V_{AX,i}}{2} \mbox{Erfc}\left[\frac{x_{th}}{\sqrt{2 V'_{DX,i}}} \right],\nonumber\\
\langle X^{'2}_B\rangle_i^{P}&=&\frac{RT (V'_{DX,i}-1)^2 x_{th}}{\sqrt{2\pi V^{'3}_{DX,i}}}\exp\left(-\frac{x_{th}^2}{2 V'_{DX,i}} \right)+\nonumber\\
&&\frac{RT(V_{BX,i}-1)^2+V_{BX,i}}{2 V'_{DXi}}\mbox{Erfc}\left[\frac{x_{th}}{\sqrt{2 V'_{DX,i}}} \right],\nonumber\\
\langle X_A X'_B\rangle_i^{P}&=&\frac{\sqrt{T}R(V'_{DX,i}-1)C_{Xi}}{\sqrt{2\pi V^{'3}_{DXi}}}\exp\left(-\frac{x_{th}^2}{2 V'_{DX,i}} \right)+\nonumber\\
& &\frac{\sqrt{T}C_{X,i} V'_{DX,i}}{2}\mbox{Erfc}\left[\frac{x_{th}}{\sqrt{2 V'_{DX,i}}} \right].
\end{eqnarray}
where $V'_{DX,i}=RV_{BX,i}+T$ is the output variance of the detected mode and
\begin{equation}
P_{S,i}=\frac{1}{2}\mbox{Erfc}\left[\frac{x_{th}}{\sqrt{2 V'_{DX,i}}} \right]
\end{equation}
is the success probability of distilling the $\i$-th constituent of the mixed state. The total probability of success $P_S$ is then given by $P_S=\sum_i p_i P_{S,i}$.

Since the first moments of the $\hP$ quadrature are vanishingly small, the $\hP$ related elements of the covariances matrix are directly given by
\begin{eqnarray}
\langle P^{2}_A\rangle^{P}&=&\frac{1}{P_S}\sum_i p_i P_{S,i}V_{AP,i},\nonumber\\
\langle P^{'2}_B\rangle^{P}&=&\frac{1}{P_S}\sum_i p_i P_{S,i}(TV_{BP,i}+R),\nonumber\\
\langle P_A P'_B\rangle^{P}&=&\frac{\sqrt{T}}{P_S}\sum_i p_i P_{S,i}C_{P,i}.\nonumber\\
\end{eqnarray}
The covariance matrix $\mathbf{CM}^{P}$ can then be constructed from these elements. This covariance matrix fully characterizes the Gaussian part of the state and thus yields the Gaussian log-negativity by using eqn. (\ref{eq_LN}).

\begin{figure}[ht]
\includegraphics[width=0.48\textwidth]{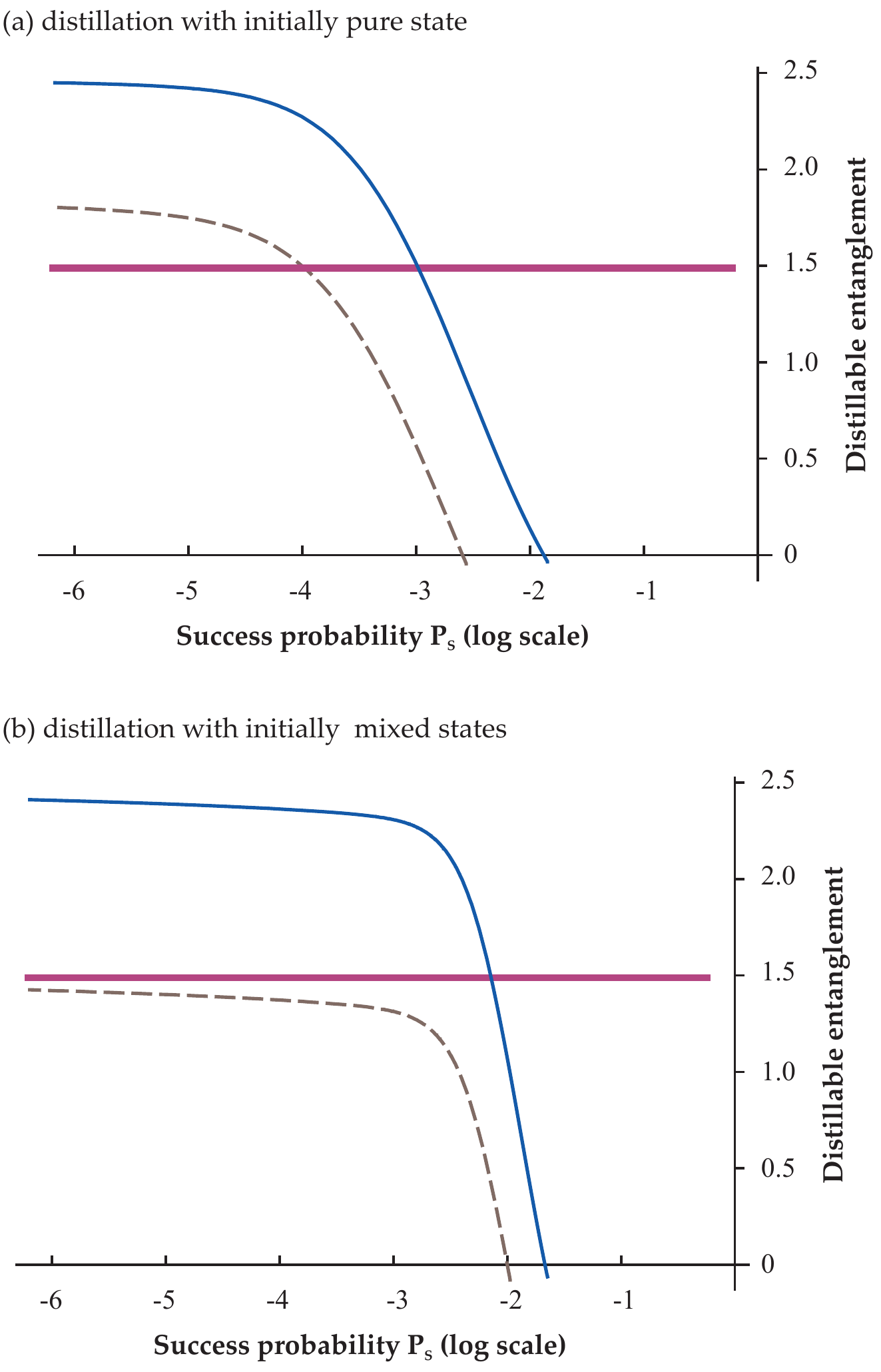}
\caption{(color online) Theoretical simulations of distillable entanglement of non-Gaussian mixed states as a function of success probability. The two plots are corresponding to two different purities of the entangled input states. In figure (a), the distillation with initially pure state is plotted ($V_A = 1/V_S$), while figure (b) shows the distillation with initially mixed states ($V_A = 1/V_S + 10$). The other parameters are taken as: $V_S = 0.1$, $V_N = 1$, $T_S = 0.5$, $T = 0.7$, $\eta_1 = 1$, $\eta_2 = 0$, $p_1 = 0.2$, $p_2 = 1 - p_1$. In both plots, the lower dashed line shows the lower bound on distillable entanglement, and the upper solid line is the upper bound on Gaussian entanglement. The bold straight line is the upper bound of the non-Gaussian distillable entanglement before distillation.}
\label{fig_theory}
\end{figure}

As we discussed in Section~\ref{sec_entevaluation}, to successfully demonstrate entanglement distillation of non-Gaussian states, the upper bound on distillable entanglement before distillation must be surpassed by the lower bound on distillable entanglement after the distillation (see also Fig.~\ref{bounds}). Due to the fragility of the lower bound, this can be only achieved for almost pure states as mentioned above. A theoretical demonstration is given in Fig.~\ref{fig_theory}. Here we consider the transmission of entanglement in a channel which is randomly blocked: The entangled state is perfectly transmitted with the probability $p_1=0.2$ and completely erased with the probability $p_2=0.8$. We assume the two squeezed states which produce entanglement have variances along the squeezed quadrature as $V_S = 0.1$, the entangling beam splitter is symmetric ($T_S=50\%$) and the tap beam splitter has a transmission of $T=0.7$. The distillation with a pure entangled state ($V_A = 1/V_S$) as well as a mixed state ($V_A = 1/V_S + 10$) are investigated as a function of the success probability, and shown in Fig.~\ref{fig_theory} (a) and (b) respectively. Following the theory of section~\ref{sec_bounds}, we calculate the upper bound on distillable entanglement of the non-Gaussian state before distillation as shown by the bold straight lines in Fig.~\ref{fig_theory}. The lower bounds on distillable entanglement after distillation are computed and shown in Fig.~\ref{fig_theory} by the dashed lines. We see that the proof of entanglement distillation of non-Gaussian states already fails for a mixed state with a small amount of excess noise. The upper bounds on {\it Gaussian} entanglement after distillation are also computed and shown in Fig.~\ref{fig_theory} by solid lines. As the success probability reduces, the Gaussian entanglement increases. Furthermore, when it surpasses the upper bound on distillable entanglement before distillation (bold solid lines) at a certain low success probability, the distilled state is Gaussified as well and thus we can justify a Gaussian state in the entanglement measure.

\section{Experimental realization}\label{experiment}

\begin{figure*}
\includegraphics[width=0.7\textwidth]{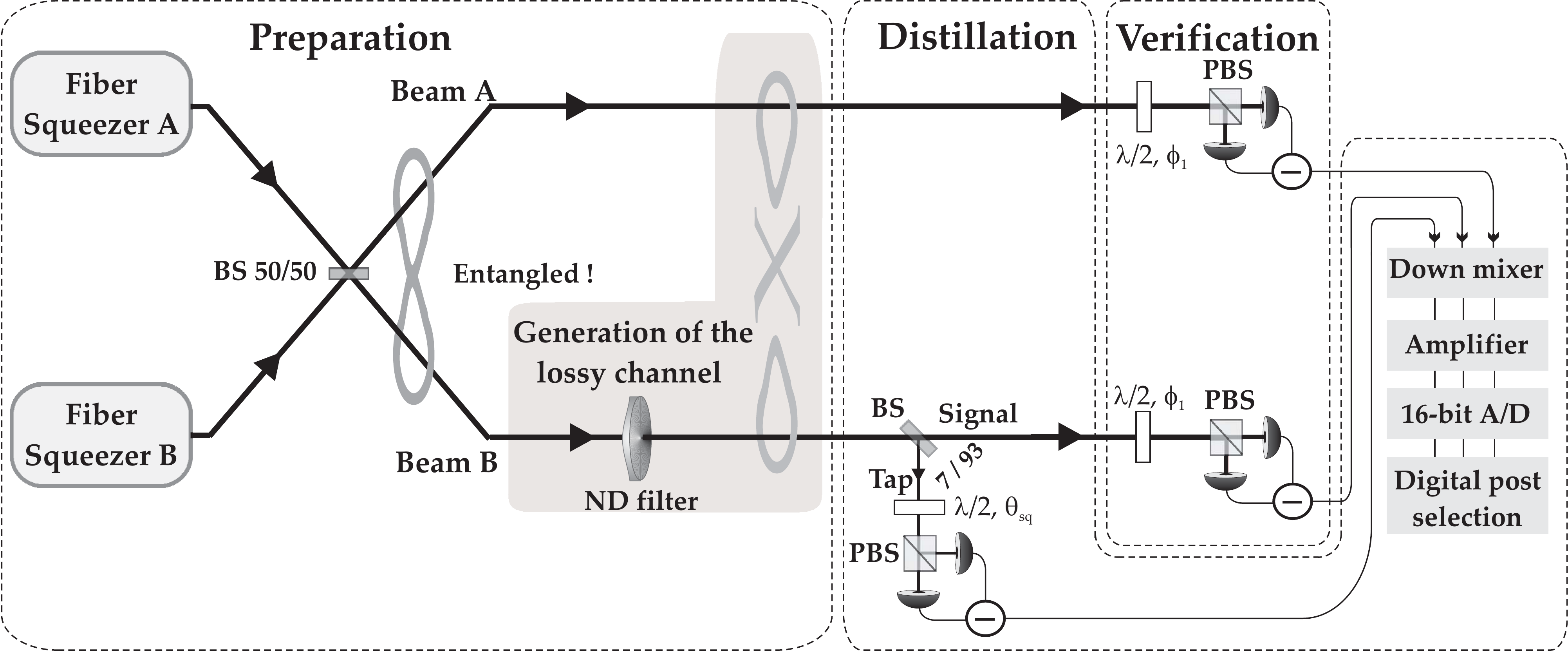}
\caption{Schematics of the experimental setup for the preparation, distillation and verification of the distillation of entanglement from a non-Gaussian mixture of polarization entangled states.}
\label{entdistsetup}
\end{figure*}

The experimental realization of the distillation of corrupted entangled states consists of three parts as schematically illustrated in Fig.~\ref{entdistsetup}: the preparation, distillation and verification. In the following we describe each part.

\subsection{Generation of polarisation squeezing and entanglement}

\begin{figure}[ht] \begin{center}
\includegraphics[width=0.48\textwidth]{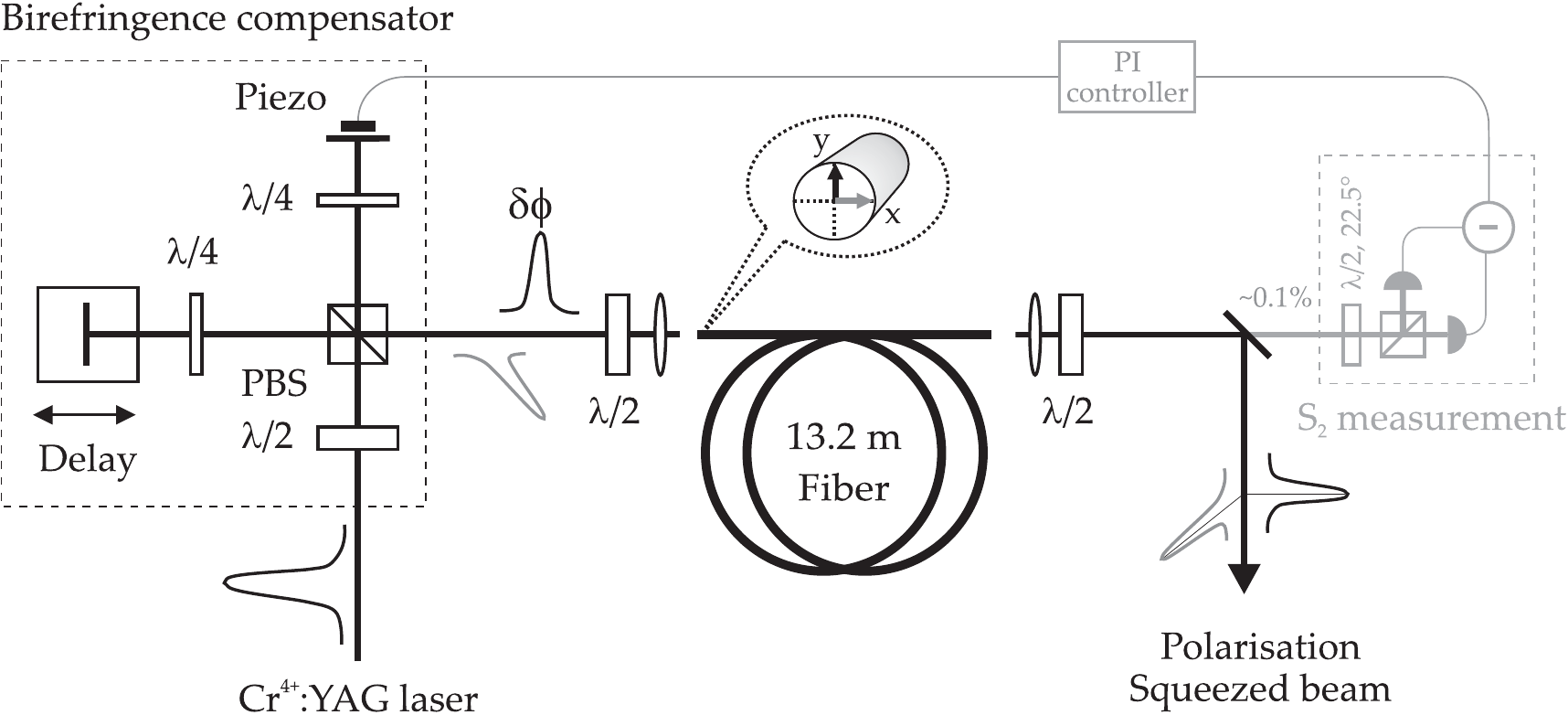}
\end{center}
\caption{Setup for the generation of polarisation squeezing. The fiber is a 13.2 meters long polarization-maintaining 3M FS-PM-7811 fiber with a mode field diameter of 5.7 $\mu$m and a beat length of 1.67 mm. The interferometer in front of the fiber introduces a phase shift $\delta \phi$ between the two orthogonally polarised pulses to pre-compensate for the birefringence. $\lambda/4$, $\lambda/2$: quarter--, half--wave plates, PBS: polarising beam splitter. PZT: Piezo-electric element.}
\label{polsqueezingsetup}
\end{figure}

The generation of polarization squeezed beams serves as the first step for the demonstration of entanglement distillation. Here we exploit the Kerr nonlinearity of silica fibers experienced by ultrashort laser pulses for the generation of quadrature squeezed states. Fig.~\ref{polsqueezingsetup} depicts the setup for the generation of a polarization squeezed beam. A pulsed (140 fs) Cr$^{4+}$:YAG laser at a wavelength of 1500~nm and a repetition rate of 163 MHz is used to pump a polarization-maintaining fiber. Two linearly polarized light pulses with identical intensities are traveling in single pass along the orthogonal polarization axes ($x$ and $y$) of the fiber. Two quadrature squeezed states, the squeezed quadrature of which are skewed by $\theta_\mathrm{sq}$ from the amplitude direction, are thereby independently generated. After the fiber the emerging pulses are overlapped with a $\pi/2$ relative phase difference. The relative phase difference is achieved using a birefringence pre-compensation, an unbalanced Michelson-like interferometer~\cite{heersink03.pra, heersink05.ol, dong08.ol, fiorentino01.pra}. This is controlled by a feedback locking loop based on a $S_2$ measurement of a small portion ($\leq$0.1\%) of the fiber output. The measured error signal is fed back to the piezo-electric element of the pre-compensation via a PI controller, so that the $S_2$ parameter of the output mode vanishes. This results in a circularly polarized beam at the fiber output ($\langle\hat{S}_1\rangle=\langle\hat{S}_2\rangle=0$, $\langle\hat{S}_3\rangle=\langle\hat{S}_0\rangle=\alpha^{2}$). The corresponding Stokes operator uncertainty relations are reduced to a single nontrivial one in the so-called $\hat{S_{1}}-\hat{S_{2}}$ dark plane: $\Delta^{2} \hat{S}_{\theta}\, \Delta^{2} \hat{S}_{\theta+\pi/2}\geq |\langle\hat{S}_3\rangle|^{2}$, where $\hat{S}(\theta) = \cos(\theta) \hat{S}_1 + \sin(\theta) \hat{S}_2$ denotes a general Stokes parameter rotated by $\theta$ in the dark $\hat{S_{1}}-\hat{S_{2}}$ plane with $\langle\hat{S}_{\theta}\rangle=0$. Therefore, polarization squeezing occurs if $\Delta^{2}\hat{S}_{\theta}<|\langle \hat{S}_{3}\rangle|=\alpha^{2}$, in which $\Delta^{2}\hat{S}_{\theta}$ can be directly measured in a Stokes measurement~\cite{heersink05.ol}. As the noise of Stokes parameters $\hat{S}_{\theta}$ is linked to the quadrature noise of the Kerr squeezed modes in the same angle ($\Delta^{2}\hat{S}_{\theta}\approx\alpha (\delta\hat{X}_{x,\theta}-\delta\hat{X}_{y,\theta})/\sqrt{2}\approx\alpha^2 \Delta^2 \hat{X}_{\theta}$~\cite{heersink05.ol}), the squeezed Stokes operator is $\hat{S}(\theta_\mathrm{sq})$ and the orthogonal, anti-squeezed Stokes operator is $\hat{S}(\theta_\mathrm{sq}+\pi/2)$. Due to the equivalence between the polarization squeezing and vacuum squeezing~\cite{josse03.prl}, we utilize the conjugate quadratures $\hat{X}$ and $\hat{P}$ to denote the polarization squeezed and anti-squeezed Stokes operators.

To generate polarization entanglement two identical polarization-maintaining fibers are used. Two polarization squeezed beams, labeled A and B, are then generated. By balancing the transmitted optical power of the two fibers, the two resultant polarization squeezed beams have identical squeezing angles, squeezing and anti-squeezing properties. The two polarization squeezed beams are then interfered on a 50/50~beam splitter (Fig.~\ref{entdistsetup}) with the interference visibility aligned to be $>98\%$. The relative phase between the two input beams is locked to $\pi/2$ so that the two output beams after the beam splitter have equal intensity and are maximally entangled. The two entangled outputs remain circularly polarised, thus the quantum correlations between them are lying in the dark $\hat{S}_1-\hat{S}_2$ plane with the signatures $\hat S_\mathrm{A}(\theta_{sq})+\hat S_\mathrm{B}(\theta_{sq})\rightarrow0$ and $\hat S_\mathrm{A}(\theta_{sq}+\pi/2)-\hat S_\mathrm{B}(\theta_{sq}+\pi/2)\rightarrow0$ (or $\hat X_\mathrm{A}+\hat X_\mathrm{B}\rightarrow0$ and $\hat P_\mathrm{A}-\hat P_\mathrm{B}\rightarrow0$).

\subsection{Preparation of a non-Gaussian mixed state}

The preparation of a non-Gaussian mixture of polarization entangled states is implemented by transmitting one of the entangled beams, e.g. beam B, through a controllable neutral density filter (ND). The filter is implemented to produce a lossy channel with $N=45$ different transmittance levels, ranging from 0.1 to 1 in steps of 0.9/44. The entangled beam is then transmitted through the lossy channel with 45 realizations. Combining all these realizations a non-Gaussian mixed state, such as the one described by eqn.~(\ref{sum}), is achieved, with the probabilities $p_i$ all being identical. However, after the measurement we can select a certain probability envelope function to give the different channels pre-specified probability weights. With this technique we can easily implement different transmission scenarios (see e.g. Fig.~\ref{discrete}-1, Fig.~\ref{continuous1}-1, and Fig.~\ref{continuous2}-1). As a result of the lossy transmission, the Gaussian entanglement between the two beams A and B are degraded or completely lost.

\subsection{Entanglement distillation}

The distillation operation consists of a measurement of $\hat{X}$ on a small portion of the mixed entangled beam. This is implemented by tapping 7\% of beam B after the ND filter using a beam splitter. The measurement is followed by a probabilistic heralding process where the remaining state is kept or discarded, conditioned on the measurement outcomes: e.g. if the outcome of the weak measurement is larger than the threshold value, $X_{th}$, then the state is kept. Note that the signal heralding process could in principle be implemented electro-optically to generate a freely propagating distilled signal state. However, to avoid such complications, our conditioning is instead based on digital data post-selection using a verification measurement on the conjugate quadratures $\hat{X}$ and $\hat{P}$ of the beams A and B.

\subsection{The tap and verification measurement}

The tap and verification measurement are accomplished simultaneously by three independent Stokes measurement apparatuses. Each measurement apparatus consists of a half-wave plate and a polarizing beam splitter (PBS). Since the light beam is circularly ($S_3$) polarized, a rotation of the half-wave plate enables the measurement of different Stokes parameters lying in the 'dark' polarization plane. For the tap measurement the half-wave plate is always set at the angle corresponding to $\hat{X}$ in the 'dark' polarization plane. Via the verification measurement setup, the Gaussian properties of the entangled states are characterized by measuring the entries of the covariance matrix. By generating near symmetric states and choosing a proper reference frame, we assume that the intra-correlations (such as $\langle \hat{X}_A \hat{P}_A \rangle$) are zero. The measurements of these entries are accomplished by applying polarization measurements of beam A and B with both the half-wave plates set to the angle corresponding to either $\hat{X}$ or $\hat{P}$ in the 'dark' polarization plane.
The outputs of the PBS are detected by identical pairs of balanced photo-detectors based on 98\% quantum efficiency InGaAs PIN-photodiodes and with an incorporated low-pass filter in order to avoid ac saturation due to the laser repetition oscillation. The detected AC photocurrents are passively pairwise subtracted and subsequently down-mixed at 17 MHz, low-pass filtered (1.9 MHz), and amplified (FEMTO DHPVA-100) before being oversampled by a 16-bit A/D card (Gage CompuScope 1610) at the rate of $10^7$ samples per second. The time series data are then low-passed with a digital top-hat filter with a bandwidth of 1 MHz. After these data processing steps, the noise statistics of the Stokes parameters are characterized at 17 MHz relative to the optical field carrier frequency ($\approx$200 THz) with a bandwidth of 1 MHz. The signal is sampled around this sideband to avoid classical noise present in the frequency band around the carrier~\cite{schiller96.prl}.
For each polarization measurement, the detected photocurrent noise of beam A and B and the tap beam were simultaneously sampled for $2.4\times10^{8}$ times, thus the self and cross correlations between the data set of A and B could thereby easily be characterized. The covariance matrix was subsequently determined and the log-negativity was calculated according to eqn.~(\ref{eq_LN}).

\section{Experimental results}\label{Experimental results}

\begin{figure}[ht]
\includegraphics[width=0.48\textwidth]{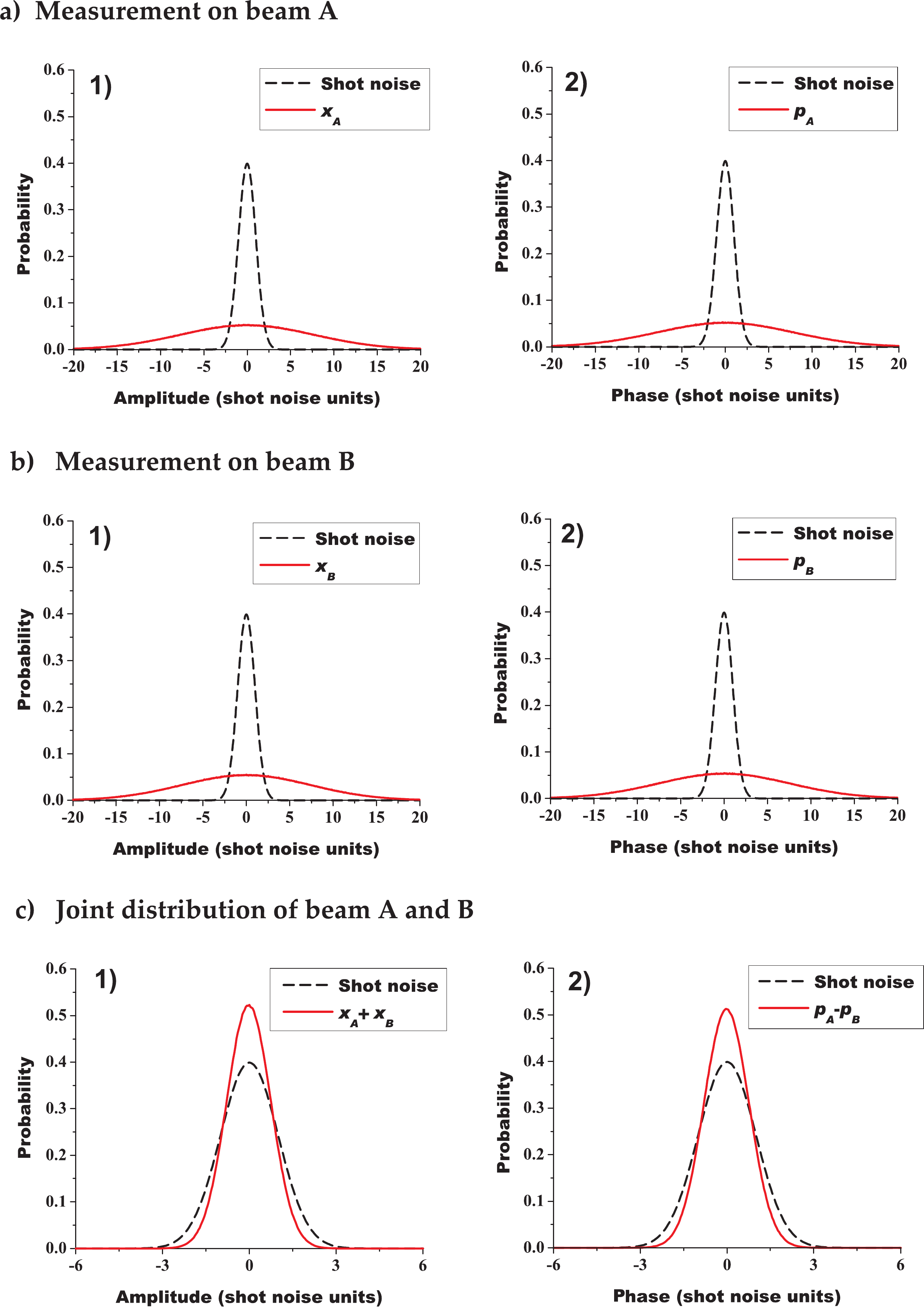}
\caption{(color online). Experimentally measured marginal distributions associated with the (a) $X$ and $P$ of beam A, (b) $X$ and $P$ of beam B and (c) the joint measurements $X_A+X_B$ and $P_A-P_B$. The black and red curves are the distributions for shot noise and the quadrature on measurement, respectively.} \label{entdistribution}
\end{figure}

For perfect transmission (corresponding to no loss in beam B), the marginal distributions of the entangled beams, A and B, along the quadrature $X$ and $P$ are plotted in Fig.~\ref{entdistribution}. In Fig.~\ref{entdistribution}(a) the procedure of realizing the different noisy channels is shown. The sampled data of different attenuation channels is concatenated according to the different weights of the transmission probabilities. These samples then provide the measurement data for the distillation procedure. From Fig.~\ref{entdistribution}(a) and Fig.~\ref{entdistribution} (b) we can see that, each individual mode exhibits a large excess noise (measured fluctuation $> 17$~dB). However the joint measurements on the entangled beams A and B exhibit less noise fluctuation than the shot noise reference, as shown in Fig.~\ref{entdistribution}(c). The observed two-mode squeezing between beam A and B is $-2.6\pm 0.3$ dB and $-2.4\pm0.3$ dB for $\hat{X}$ and $\hat{P}$, respectively. From the determined covariance matrix we compute the log-negativity to be $0.76\pm0.08$.

To experimentally demonstrate the distillation of entanglement out of non-Gaussian noise, three different lossy channels are considered: the discrete erasure channel, where the transmission randomly alternates between two different levels, and two semi-continuous channels, where the transmittance alternates between 45 different levels with specified probability amplitudes. The probability distributions of the transmittance for the discrete channel and the continuous channels are shown in Fig.~\ref{discrete}-1, Fig.~\ref{continuous1}-1, and Fig.~\ref{continuous2}-1, respectively.

\subsection{The discrete lossy channel}

\begin{figure}[ht]
\includegraphics[width=0.48\textwidth]{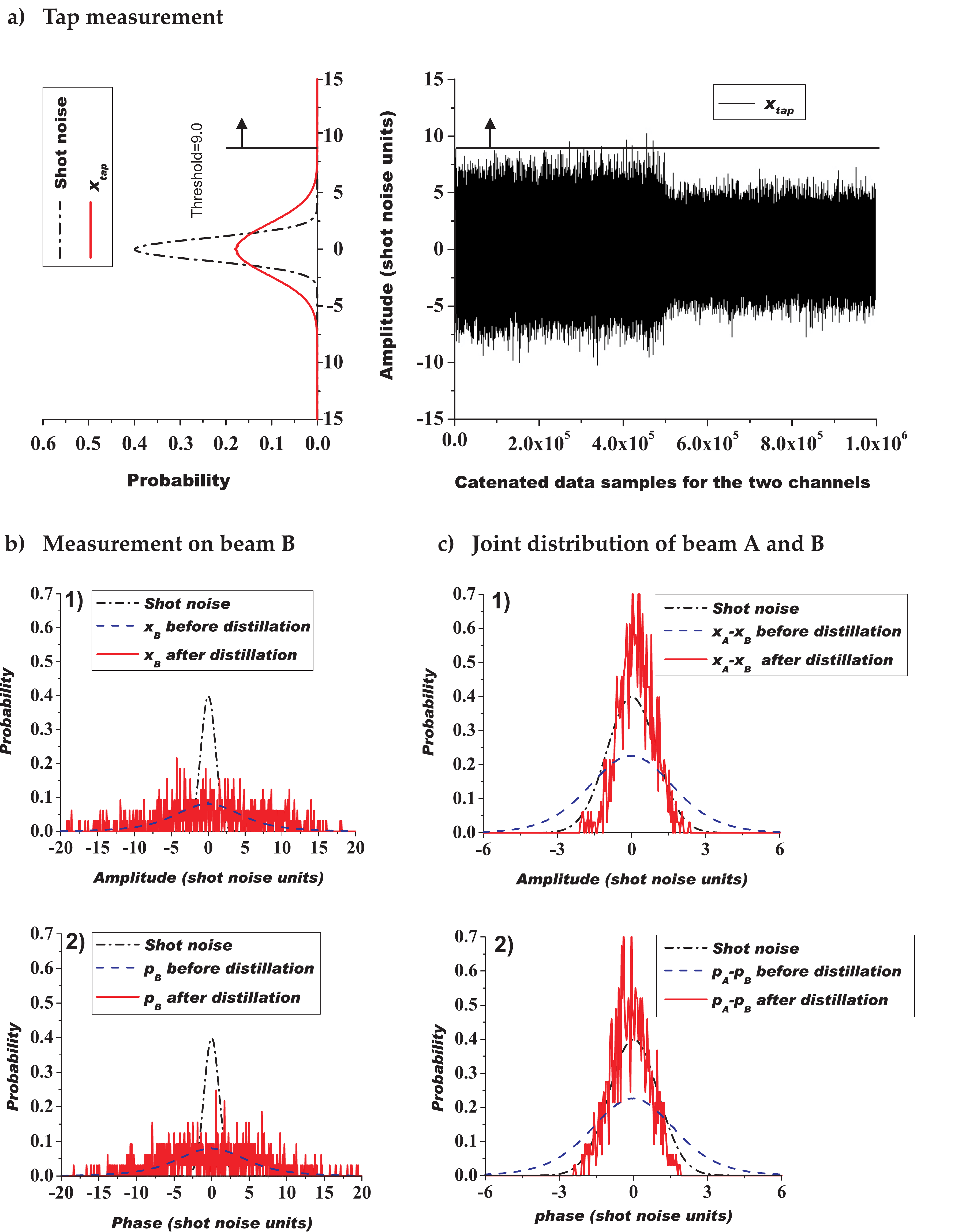}
\caption{(color online). Experimentally measured marginal distributions illustrating the effect of distillation. (a) Example of concatenated sampled data and the resulting marginal
distribution for the amplitude quadrature in the tap measurement. The vertical line indicates the threshold value chosen for this realization. (b) Marginal distributions associated with the measurements of $X$ and $P$ of beam B (two left figures) and (c) the joint measurements $X_A+X_B$ and $P_A-P_B$ (two right figures). The black, blue and red curves are the distributions for shot noise, the mixed state before distillation and after distillation, respectively. Inset: phase-space representation of the non-Gaussian mixed state and the post-selection procedure used in the measurements. The black vertical line indicates the threshold value.} \label{entdistribution1}
\end{figure}

The discrete erasure channel alternates between full (100\%) transmission and 25\% transmission at a probability of $0.5$. Each realisation is concatenated to each other with identical weights. The concatenation procedure yields the same statistical values as true randomly varying data. After transmission the resulting state is a mixture of a highly entangled state and a weakly entangled state. In the inset of Fig.~\ref{entdistribution1}, we show marginal distributions illustrating the single beam statistics of the individual components of the mixture. The statistics of beam B is seen to be contaminated with the attenuated entangled state thus producing non-Gaussian statistics. For this state we measure the correlations in $\hat{X}$ and $\hat{P}$ to be above the shot noise level by $5.5\pm 0.3$ dB and $5.6\pm 0.3$ dB, respectively, and the Gaussian LN to be $-1.63\pm 0.02$. The Gaussian entanglement is completely lost as a result of the introduction of such time-dependent loss. This is in stark contrast to the scenario where only stationary loss (corresponding to Gaussian loss) is inflicting the entangled states. In that case, a certain degree of Gaussian entanglement will always survive, although it will be small for high loss levels.

The state is then fed into the distiller and we perform homodyne measurements on beam A, beam B and the tap beam simultaneously. By measuring $\hat{X}$ in the tap we construct the distribution shown by the red curve in the left hand side of Fig.~\ref{entdistribution1}(a). The data trace of the mixed tap signal is plotted accordingly on the right hand side. The measurements of $\hat{X}$ and $\hat{P}$ of the signal entangled states were recorded as well. For simplicity, we only show the distribution for the beam B (in Fig.~\ref{entdistribution1}(b-1),(b-2)) and the joint distribution of beam A and B (in Fig.~\ref{entdistribution1}(c-1),(c-2)). The blue (dashed) and red curves denote the distributions before and after the post-selection process, respectively. From the blue curves shown in Fig.~\ref{entdistribution1}, we can see that the entanglement between A and B is lost due to the non-Gaussian noise. Performing postselection on this data by conditioning it on the tap measurement outcome (denoted by $X_{th}=9.0$), we observe a recovery of the entanglement. That is, the correlated distribution of the signal turns out to be narrower than that of the shot noise (as shown by red curves in Fig.~\ref{entdistribution1}(c).

\begin{figure}[ht]
\includegraphics[width=0.5\textwidth]{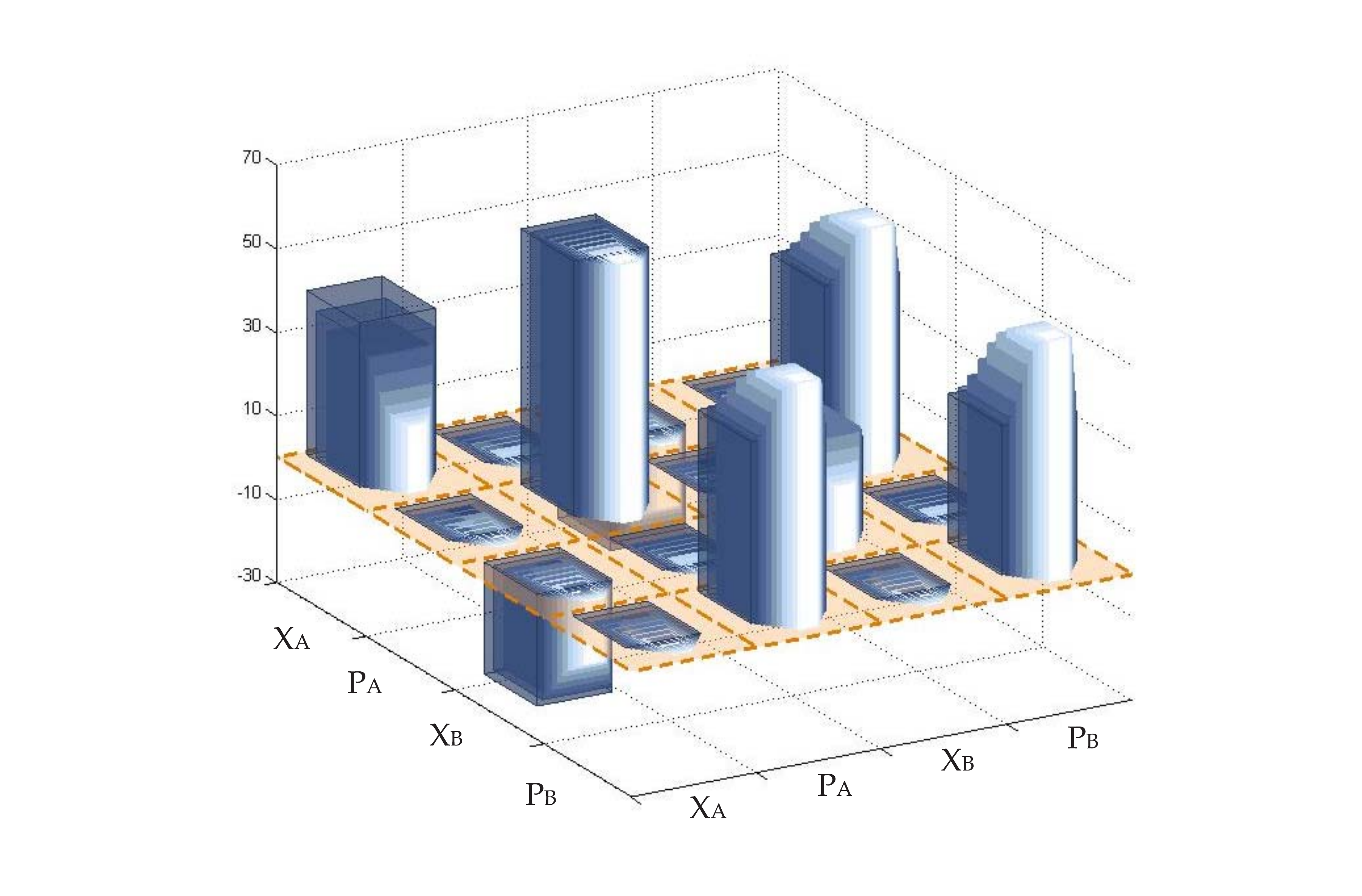}
\caption{(color online). Reconstructed covariance matrices of distilled entangled states. The brown segmented plane shows the region for the individual elements in the covariance matrix. The sub-bars represent the results of our distillation protocol for 10 different threshold values postseletion threshold values from $X_{th}=0.0$ to $X_{th}=9.0$ with a step of 1.0.} \label{barplot_covariance}
\end{figure}

Using the data shown in Fig.~\ref{entdistribution1}, a tomographic reconstruction of the covariance matrices of the distilled entangled state was carried out. 
From these data we determined the most significant eight of the ten independent parameters of the 	
covariance matrix, namely the variances of four quadratures $\hat{X}_A$,$\hat{X}_B$, $\hat{P}_A$, $\hat{P}_B$ and covariances between all pairs of quadratures of the entangled beams A and B. As mentioned before, the intra-correlations were ignored. The resulting covariance matrices are plotted in Fig.~\ref{barplot_covariance} for ten different postseletion threshold values from $X_{th}=0.0$ to $X_{th}=9.0$ with a step of 1.0. With increasing postselection threshold the distillation becomes stronger, as shown by the reduction of the quadrature variances of $\hat{X}_A$, $\hat{X}_B$ and the increase of the quadrature variances of $\hat{P}_A$, $\hat{P}_B$. Moreover, the reduction (or increase) of the covariances $C(\hat{X}_A,\hat{X}_B)$ ($C(\hat{P}_A,\hat{P}_B)$) was shown slightly slower. Consequently, the entanglement of the two modes A and B was enhanced 	
by the distillation. 	

\begin{figure}[ht]
\includegraphics[width=0.48\textwidth]{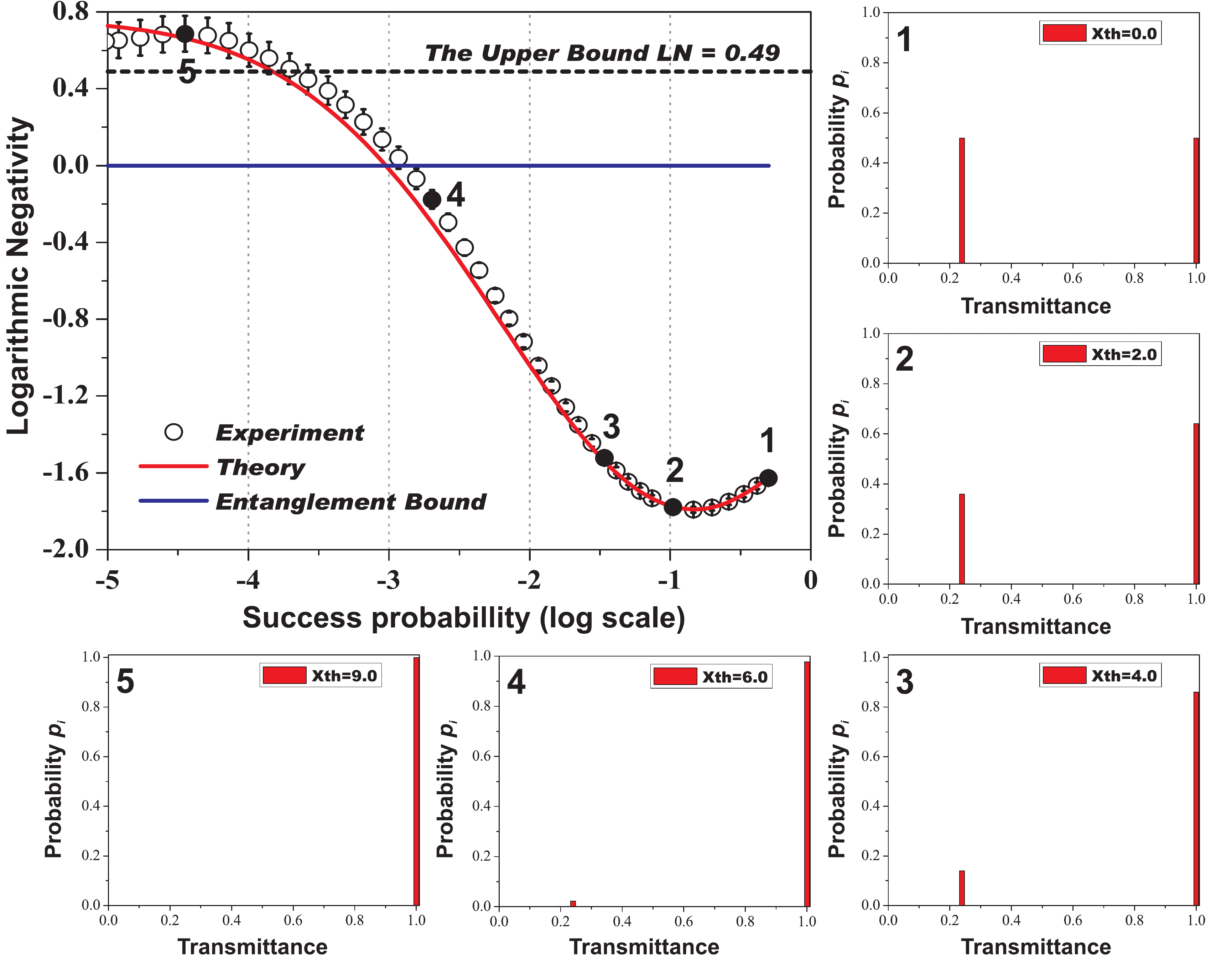}
\caption{(color online). Experimental and theoretical results outlining the distillation of an entangled state from a discrete lossy channel. The experimental results are marked by circles and the theoretical prediction is plotted by the red solid line. The bound for Gaussian entanglement is given by the blue line, and the upper bound for total entanglement before distillation is given by the black dashed line. Both bounds are surpassed by the experimental data. The weight of the two constituents in the mixed state after distillation for various threshold values is also experimentally investigated and shown in the plots labeled by 1-5. The error bars of the log-negativity represent the standard deviations.} \label{discrete}
\end{figure}

Furthermore, the distilled entanglement, or log-negativity, was investigated as a function of the success probability, as shown in Fig.~\ref{discrete} by black open circles. The error bars of the distilled log-negativity depend on two  contributions: First the measurement error, which is mainly associated with the finite resolution of the A/D converter and noise of the electronic amplifiers. This is considered by estimating the experimental error for all the elements of the covariance matrix as '0.03'. The measurement error for the LN can be simulated by a Monte-Carlo model. Second, the statistical error is due to the finite measurement time and the postselection process. It is considered by adding a scaled term $\sqrt{2/(N-1)}$, where $N$ denotes the number of postselected data~\cite{frieden83.book}. The probability distributions of the two superimposed states in the mixture after distillation are shown for different postselection thresholds, corresponding to $X_{th}=0.0, 2.0, 4.0, 6.0, 9.0$, labeled by 1-5 in order. The plots explicitly show the effect of the distillation protocol, when the postselection threshold increases, the Gaussian LN increases, ultimately approaching the LN of the input entanglement without losses. The probability distribution tends to a single valued distribution, therefore the mixture of the two Gaussian entangled states reduces to a single highly entangled Gaussian state, thus demonstrating the act of Gaussification. However, the amount of distilled data, or success probability, decreases, causing an increase in the statistical error on the distillable entanglement. Based on the experimental parameters, a theoretical simulation is plotted by the red curve and shows a very good agreement with the experimental results.

To further investigate whether the total entanglement is increased after distillation, we compute the upper bound for the LN before distillation and verify that this bound can be surpassed by the Gaussian LN after distillation. The upper bound of LN without the Gaussian approximation is computable from the LN of each Gaussian state in the mixture~\cite{vidal02.pra}, and we find $LN_{upper}=0.49$, which is shown in Fig.~\ref{discrete} by the dashed black line. We see that for a success probability around $10^{-4}$ the Gaussian LN crosses the upper bound for entanglement. Since the state at this point is perfectly Gaussified we may conclude that the total entanglement of the state has indeed increased as a result of the distillation. Fig.~\ref{discrete}-5 gives another explicit explanation by showing that the probability contribution from the 75\% attenuated data reaches 0 when the post selection threshold is set to $X_{th}=9.0$, which corresponds to the distilled entanglement of $LN^{P}_{S}=0.67\pm0.08$ with a success probability $P_S=1.69\times10^{-5}$. On the other hand, from Fig.~\ref{discrete}-3 and Fig.~\ref{discrete}-4, we see that even a small contribution from the 75\% attenuated data will reduce the useful entanglement for Gaussian operations.

\subsection{The continuous lossy channel}

\begin{figure}[th]
\includegraphics[width=0.48\textwidth]{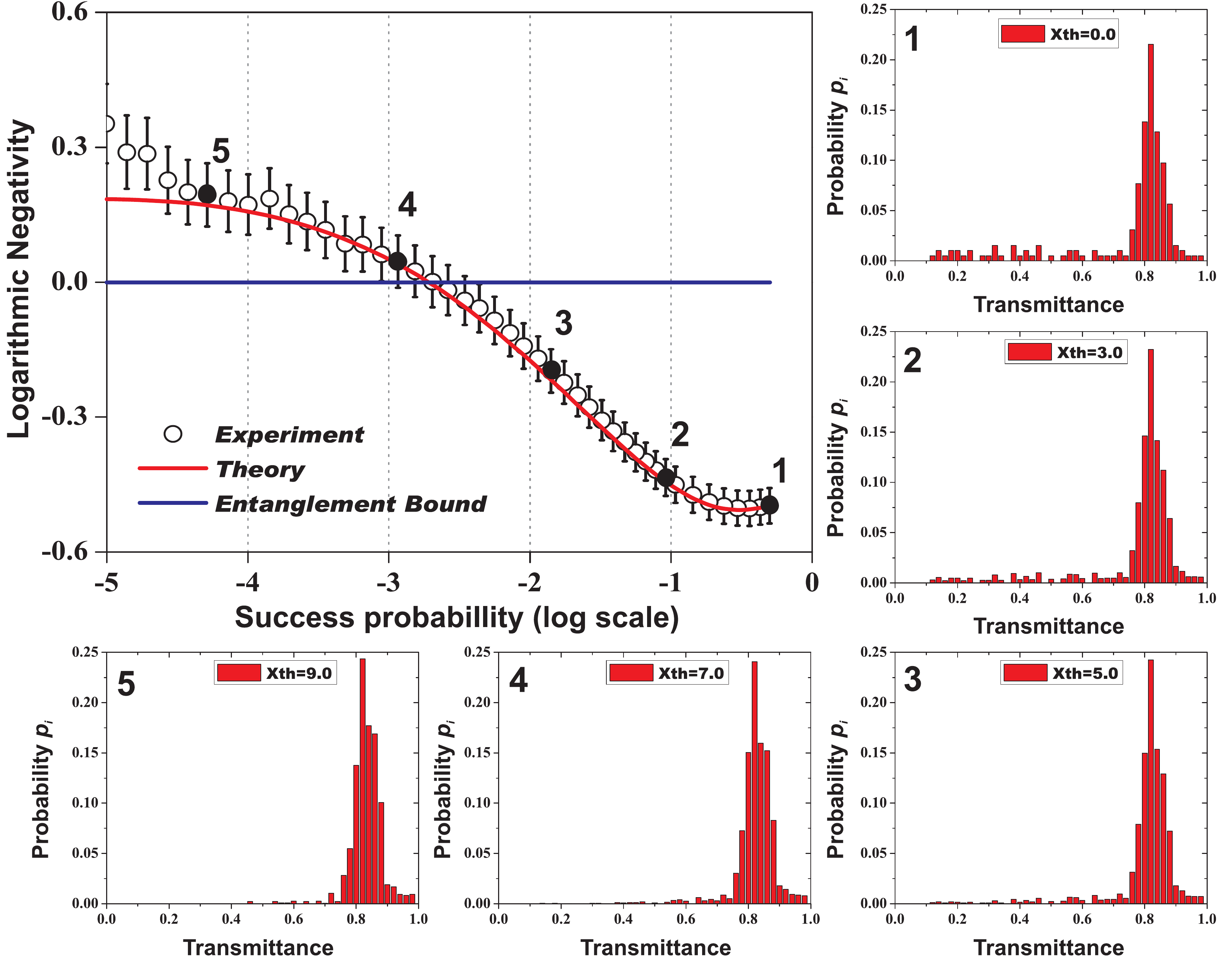}
\caption{(color online). Experimental and theoretical results outlining the distillation of an entangled state from a simulated continuous lossy channel in which the peak of the transmittance distribution is displaced from 1 to 0.8. The experimental results are marked by circles and the theoretical prediction by the red solid curve. The evolution of the weights of the various constituents in the mixed state as the threshold value is changed is shown in the figures labeled 1-5. The error bars of the log-negativity represent the standard deviations.} \label{continuous1}
\end{figure}

\begin{figure}[th]
\includegraphics[width=0.48\textwidth]{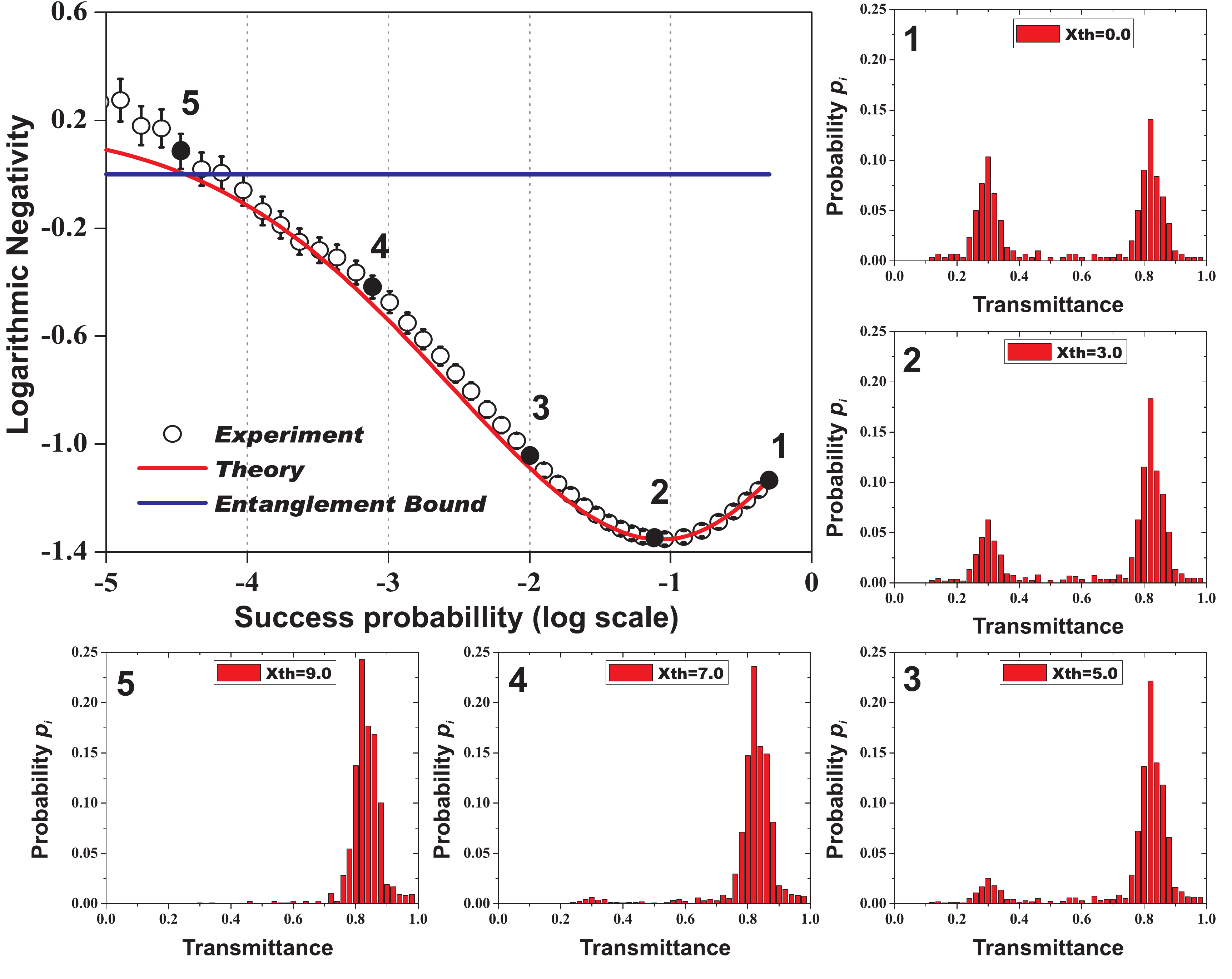}
\caption{(color online). Experimental and theoretical results outlining the distillation of an entangled state from a simulated continuous lossy channel in which the transmittance levels are distributed as such that there are two peaks at both high transmittance levels (0.8) and low transmittance levels (0.3). The experimental results are marked by circles and the theoretical prediction by the red solid curve. The evolution of the weights of the various constituents in the mixed state as the threshold value is changed is shown in the figures labeled 1-5. The error bars of the log-negativity represent standard deviations.} \label{continuous2}
\end{figure}

We now generalize the lossy channels to have a continuously transmittance distribution. The channel transmittance distribution is simulated by taking 45 different transmission levels as opposed to the two levels in the previous section. In Ref.~\cite{dong08.nat} we reported a channel whose transmittance is given by an exponentially decaying function with a long tail of low transmittances, which simulates a short-term free-space optical communications channel where atmospheric turbulence causes scattering and beam pointing noise~\cite{book}. We showed that the entanglement available for Gaussian operations can be successfully distilled from $-0.11\pm0.05$ to $0.39\pm0.07$ with a success probability of $1.66\times10^{-5}$. However, in practical scenarios for a transmission channel, the highest transmittance level may not have the biggest weight in the probability distribution and therefore the distributed peak may be displaced from the 100\% transmittance level. Further, there might be more than one peak in the probability distribution diagram. For instance, due to some strong beam pointing noise another distributed peak will appear in the area of low transmittance levels. In the following we will test the performance of the distillation protocol for two different transmittance distributions. First, when the mixed state has a peak of the transmittance distribution which is displaced from 1 to 0.8 (Fig.~\ref{continuous1}-1). Second, when we incorporate a second peak which is located around the transmittance level of 0.3 (Fig.~\ref{continuous2}-1).

As shown in Fig.~\ref{continuous1}, after propagation through the one-peak displaced channel the Gaussian LN of the mixed state is found to be $-0.50\pm0.04$, which is below the bound for available entanglement
(shown by the solid blue line) and substantially lower than the original value of $0.76\pm0.08$. The state is subsequently distilled and the change in the Gaussian LN as the threshold value increases (and the success probability decreases) was investigated both experimentally (black open circles) and theoretically (red curve). The evolution of the mixture is directly visualized in the series of probability distributions in Fig.~\ref{continuous1}-1 to~\ref{continuous1}-5 corresponding to the postselection thresholds $X_{th}$=0.0, 3.0, 5.0, 7.0, 9.0 respectively. We see that the distribution weights of the low transmittance levels is gradually reduced,  while the weights of the high transmittance levels is increased as the post-selection process becomes more and more restrictive by increasing the threshold value. E.g. for $X_{th}=9$ the probabilities associated with transmission levels lower than 0.7 are decreased from 20\% before distillation to 1.4\% and the probability for transmission levels higher than 0.7 transmission are increased to 98.6\% as opposed to 80\% before distillation. It is thus clear from these figures that the highly entangled states in the mixture have larger weight after distillation, and the corresponding Gaussian LN after distillation rises to $0.19\pm0.06$ with the success probability of $5.16\times10^{-5}$.

We now turn to investigate the distillation after propagation through the two-peak displaced channel as shown in Fig.~\ref{continuous2}-1. Before distillation the Gaussian LN of the mixed state is found to be $-1.13\pm0.02$. Likewise, the relation between the distilled Gaussian LN and the success probability was investigated both experimentally and theoretically. The results are shown in Fig.~\ref{continuous2} by black open circles and the red curve, respectively. Through the probability distribution plots in Fig.~\ref{continuous2}-1 to~\ref{continuous2}-5, the evolution of the mixture corresponding to different choices of postselection thresholds ($X_{th}$=0.0, 3.0, 5.0, 7.0, 9.0 respectively) was illustrated with the same trend that we see on the distillation after the one-peak displaced channel. For $X_{th}=9$ the probabilities associated with transmission levels lower than 0.7 are decreased from 48\% before distillation to 1.6\% and the probability for transmission levels higher than 0.7 transmission are increased to 98.4\% as opposed to 52\% before distillation, and the corresponding Gaussian LN after distillation reaches $0.19\pm0.06$ with the success probability of $3.39\times10^{-5}$.

After having shown the successful entanglement distillation on different distributions of non-Gaussian noise, we should note that the successful entanglement distillation depends on the transmittance distribution of the lossy channel. For some distributions, the success probability for distilling available entanglement for Gaussian operations will be extremely small or not be possible. For example, after a channel with the transmittance uniformally distributed, the Gaussian log-negativity $LN_{S}=-1.26\pm0.02$ before distillation will only be increased to $-0.76\pm0.03$ with a success probability of $1.32\times10^{-5}$. In general more uniform transmittance distributions turned out to be more difficult for the distillation procedure. Distributions with high probabilities for high transmission levels and pronounced tails and peaks at low transmission levels (as would be expected in atmospheric channels) are more suited.

\section{Summary}

In summary, we have proposed a simple method of distilling entanglement from single copies of quantum states that have
undergone attenuation in a lossy channel with varying transmission. Simply by implementing a weak measurement based on a beam splitter and a homodyne detector, it is possible to distill a set of highly entangled states from a larger set of unentangled states if the mixed state is non-Gaussian. The protocol was successfully demonstrated for a discrete erasure channel where the transmittance alternates between 2 levels and two semi-continuous transmission channels where the transmission levels span 45 levels with specified distributions, respectively. We show that the degree of Gaussian entanglement (which is relevant for Gaussian information processing) was substantially increased by the action of distillation. Moreover, we proved experimentally that the total entanglement was indeed increased for the discrete channel. We found that the successful entanglement distillation depends on the transmittance distribution of the lossy channel.
The demonstration of a distillation protocol for non-Gaussian noise provides a crucial step towards the construction of a quantum repeater for transmitting continuous variables quantum states over long distances in channels inflicted by non-Gaussian noise.

\subsection*{Acknowledgments}

This work was supported by the EU project COMPAS (no.212008), the Deutsche Forschungsgesellschaft and the Danish Agency for Science Technology and Innovation (no. 274-07-0509). ML acknowledges support from the Alexander von Humboldt Foundation and RF acknowledges MSM 6198959213, LC 06007 of Czech Ministry of Education and 202/07/J040 of GACR.

\end{document}